# The Optical Gravitational Lensing Experiment.
## Variable stars in the Sculptor dwarf spheroidal galaxy [1]

J. Kaluzny, M. Kubiak, M. Szymański, A. Udalski

Warsaw University Observatory, Al. Ujazdowskie 4, 00-478 Warsaw, Poland

e-mail: jka,mk,msz,udalski)@sirius.astrouw.edu.pl

and

W. Krzemiński

Carnegie Observatories, Las Campanas Observatory, Casilla 601, LaSerena, Chile

e-mail: wojtek@roses.ctio.noao.edu

and

M. Mateo

Department of Astronomy, University of Michigan, 821 Dennison Bldg.,
Ann Arbor, MI 48109-1090, USA, e-mail: mateo@astro.lsa.umich.edu

## Abstract

The central area of the Sculptor dwarf galaxy was surveyed for variable stars as a side-program of the OGLE project. Light curves in the V band were obtained for 226 RR Lyr stars and for 3 anomalous cepheids. One previously unknown anomalous cepheid was identified. We discovered also two variables located at the tip of the red giant branch of Sculptor. Out of 226 RR Lyr variables 135 were classified as RRab, 88 as RRc and 2 as RRd. Distribution of periods for RRab stars shows a sharp cut-off at $P = 0.475$ days. This implies that the bulk of Sculptor RR Lyr stars has metallicity [Fe/H] $\leq -1.7$ on the Zinn-West scale. The average V magnitudes of RRab variables are correlated with their periods. This effect is most probably caused by the spread of metallicities exhibited by variables. Based on the average V magnitudes of RR Lyr stars the apparent distance modulus of Sculptor was determined at $(m - M)_V = 19.71$. The color-magnitude diagram of Sculptor reaching $V \approx 21.4$ and $I \approx 20.6$ is presented. The observed width of the upper part of the red giant branch indicates range of metallicities $-2.2 \leq$ [Fe/H] $\leq -1.6$ for the Sculptor giants. The observed distribution of stars along the horizontal branch and average value of metallicity imply that age of Sculptor is similar to ages of the relatively young globular clusters from the outer galactic halo. The data about RR Lyr variables in four nearby dwarf galaxies are summarized and discussed briefly.

## 1. Introduction

The Optical Gravitational Lensing Experiment (OGLE) is a long term project with the main goal of searching for dark matter in our Galaxy by identifying microlensing events toward the Galactic Bulge (Udalski et al. 1992, 1994a). At times the Bulge is unobservable we conduct other long-term photometric programs. A complete list of side-projects attempted by the OGLE team can be found in Paczynski et al. (1995). In this paper we present study of variable stars in the Sculptor dwarf galaxy.

The Sculptor galaxy is one of satellites of the Milky Way. It was discovered by Shapley (1938)

---

as the first dwarf galaxy ever identified. The pioneering search for variable stars in Sculptor was conducted by Baade and Hubble (1939). They discovered about 40 variables, of which all but two brightest were RR Lyr stars. Thackeray (1950) reported 237 variables while van Agt (1978) published list of 602 variables and listed periods for 62 RR Lyr variables and for 3 anomalous cepheids.

## 2. Observations and data reduction

The OGLE project is conducted using the 1-m Swope telescope at Las Campanas Observatory which is operated by Carnegie Institution of Washington. A single $2048 \times 2048$ pixels Loral CCD chip, giving the scale of 0.435 arcsec/pixel is used as the detector. The initial processing of the raw frames is done automatically in near-real time. Details of the standard OGLE reduction techniques are described by Udalski et al. (1992).

The data for this project were obtained during 1992 and 1993 observing seasons. Detailed logs of observations can be found in Udalski et al. (1992, 1994b). One field centered approximately on the center of the galaxy was observed. Most of the data were obtained during the period from Jun 24 to Sep 5, 1993. The exposure times ranged from 700 to 1100 seconds with 900 seconds being the most frequent value. A total of 87 exposures in the V band and 11 exposures in the I band were collected. The seeing ranged from 1.2" to 2.2" with the average value of about 1.5".

Extraction of photometry was conducted using procedure similar to that described by Udalski et al. (1992). The DoPHOT photometry program (Schechter et al. 1993) was used to derive profile-fitting photometry. We used the DoPHOT program in the fixed-position mode. The stellar positions were provided from a list of the output photometry taken from the reduction of a "template" image. The frame mr4741 (see Udalski et al. 1994b) was chosen as a template. To cope with the effects caused by the positional changes of the point spread function, each analyzed frame was divided into a $4 \times 4$ grid of overlapping sub-frames. The point spread function showed only very small variability in these $540 \times 540$ pixel sub-images. Photometry derived for the "template" sub-frames was transformed to the common instrumental system by application of additive corrections. These corrections were equal to the aperture corrections derived for each sub-frame. The aperture corrections were calculated using the Daophot program (Stetson 1987, 1991). Subsequently instrumental photometry derived for a given sub-frame of a given frame was tied to the common instrumental system of the "template" image. Finally the data base containing photometry from all reduced frames was constructed. The actual procedure of constructing the data base was similar to procedure described by Udalski et al. (1992) and Szymański & Udalski (1993).

The data were calibrated using observations collected during the photometric night of July 13, 1993. Five fields containing 17 standard stars from the Landolt (1992) list were observed on this night. The following relations were obtained:

$$v = \text{const} + V - 0.028 \times (V - I) + 0.086 \times (X - 1.25) \tag{1}$$

$$v - i = \text{const} + 0.952 \times (V - I) + 0.056 \times (X - 1.25) \tag{2}$$

where X is the airmass and a lower case letters correspond to the instrumental magnitudes. This transformation is very similar to the transformation obtained for the 1992 season by Udalski et al. (1992). Two consecutive frames of the Sculptor obtained on the night of July 13, 1993 (frames mr4528 and mr4529 in Udalski et al. 1994b) were used to obtain VI photometry for stars from the monitored field. This photometry was subsequently used to transform instrumental photometry from the data base to the V system. A simplified transformation in the form $V = v + dv$ was adopted. To determine value of offset $dv$ an average value of (V-v) was calculated for stars with 19<V<20 and $0.9 < V - I < 1.0$. The color term of the V transformation equals 0.028 (see eq. 1). Hence, neglection of this term leads to some systematic errors which do not exceed 0.028 mag for stars with $0 < V - I < 2.0$. Colors of Sculptor RR Lyr stars discussed in the next section range from $V - I \approx 0$ to $V - I \approx 0.6$. This range includes variability of color with phase.

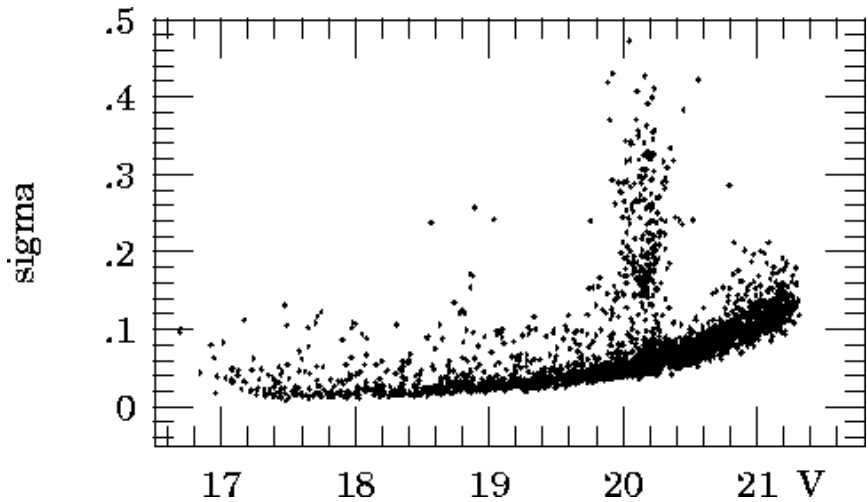

Figure 1: Standard deviation vs. average V magnitude for stars with at least 50 reliable measurements. The group of stars with large values of $rms$ present at V $\approx$ 20.1 are RR Lyr variables.

## 3.  Variable stars

A search for variable stars was made using a $\chi 2$ test. The total number of stars contained in the data base for the V filter was equal to 6196. Only light curves of 3218 stars which were measured on at least 50 frames were tested. In total, 326 objects satisfying condition $P(\chi 2) < 1 \times 10^{-4}$ were selected for further examination. The search for variables was limited to the V photometry because of the relatively low number of measurements for the I band. Of 326 candidate variables 95 turned out to be spurious or irregular variables. Almost all certain variables are RR Lyr stars. The 6 exceptions are 3 anomalous cepheids and 3 long period variables. In Fig. 1 we plotted $rms$ vs. average V magnitude for stars whose light curves were examined for variability.

### 3.1.  RR Lyr variables

The light curve parameters derived for RR Lyr variables are given in Table 1. This table contains also rectangular coordinates of variables. The ID numbers assigned to variables correspond to their running numbers in the data base for the V filter. Periods were determined using Clean (Roberts et al. 1987) and AoV (Schwarzenberg-Czerny 1991) algorithms. In all but very few cases unambiguous value of period could be derived. V1168 and V4780 are double-mode pulsators belonging to RRd variables. For RRd stars we printed the first-overtone periods in Table 1. The fundamental periods of RRd stars V1168 and V4780 were determined at 0.5415 and 0.6441 days, respectively. $N_V$ is the total number of measurements accepted for a given star (some points with very large formal errors of the photometry were rejected). Columns 4-5 are magnitudes at the maximum and minimum of brightness, respectively. The full amplitude of a light curve is given in column 6. Intensity-mean magnitudes are listed in column 7. They were calculated by numerical integration of the phased light curves. Columns 8-9 are rectangular coordinates corresponding to positions of stars on the template image. The template image (FITS format file) was submitted to the editor of A&A together with the paper (see Appendix A). Assigned types of variable are listed in column 10. A question mark in column 10 indicates uncertain classification of a given star while capital letter B indicates variable period or amplitude (i.e. Blazhko effect). The equatorial coordinates of all variables are given in Table 2. A transformation from the rectangular to the equatorial coordinates was derived based on positions of 12 stars from the Guide Star Catalogue (Lasker et al. 1988). The GSC stars were distributed more or less uniformly over the observed field. The adopted plate solution reproduces their equatorial coordinates listed in the GSC with residuals not exceeding 0.7".

Figure 2: Phased V-light curves of Sculptor RR Lyr stars. Inserted labels give variables ID and their periods.

An attempt was made to cross-correlate objects from Table 1 with the list of variables given by van Agt (1978). Unfortunately it turned out to be impossible to obtain accurate transformation between van Agt's coordinates and our system of rectangular coordinates. It seems that for many stars from van Agt's list errors of coordinates are much larger than the quoted uncertainty ±4 arcsec. No attempt was made to cross-identify variables from both lists by hand. We determined however, that there were 213 van Agt's variables inside the field monitored by us. We identified 5 randomly selected variables from the van Agt's list - namely stars V4, V103, V195, V458 V475 - which were not present on our list of variables. Their light curves were extracted from the data base and it turned out that all five stars do not show any evidence for variability. This indicates that some fraction of stars identified by van Agt (1978) as RR Lyr variables are in fact non-variable objects.

The phased light curves of Sculptor RR Lyr stars are shown in Fig. 2. The tabular data with photometry of all variables were submitted to the editor of A&A (see Appendix A). In Fig. 3 we present a period−V amplitude diagram for the Sculptor RR Lyr stars. Periods and amplitudes of RRab variables are correlated, stars with shorter periods tend to have larger amplitudes. The same relation was observed in some globular clusters harboring rich populations of RR Lyr variables. No correlation between $P$ and $A_V$ is observed for the Sculptor RRc variables. The histogram showing distribution of periods for RRab and RRc variables is shown in Fig. 4.

It is known that properties of RR Lyr stars are correlated with metallicity of their parent population. With increasing metallicity the average period $< P_{ab} >$ decreases and the blue edge of the instability strip moves toward bluer colors. Also the relative frequency of RRc stars − $N_c/(N_c + N_{ab})$ − has tendency to be higher for populations of lower metallicity. Galactic globular clusters are conventionally classified into two groups: Oosterhoff type I (metal-intermediate objects with $< P_{ab} > \approx 0.55$ days) and Oosterhoff type II (metal-poor objects with $< P_{ab} > \approx 0.65$). Recent extensive reviews of this subject and summary of observational data can be found in Sandage (1993a) and in Bono et al. (1994). For the Sculptor variables we obtained $< P_{ab} > = 0.585$ ($< \log P_{ab} > = -0.236$) and $N_c/(N_c + N_{ab}) = 88/222 = 0.40$ what corresponds formally to "Oosterhoff-intermediate" type of population . In fact no galactic globular cluster is known with $< \log P_{ab} >$ in the range of −0.21 to −0.24, although such "Oosterhoff-intermediate" clusters are known in the Large Magellanic Cloud (e.g. Bono et al. 1994). It is known that giants in the Sculptor galaxy show a range of metallicities (Kunkel and Demers 1977, Da Costa 1984). Distribution of periods of the Sculptor RRab variables shows a sharp cut-off at $P_{ab} = 0.475$ days ($\log P_{ab} = -0.32$). Sandage (1993a) obtained the following relation between the period and the metallicity at the blue fundamental edge of the instability strip:

$$\log P_{ab} = -0.122 \times [Fe/H] - 0.500 \qquad (3)$$

where [Fe/H] is on the Butler-Blanco (Butler 1975, Blanco 1992) scale. For $P_{ab} = 0.475$ days Eq. 3 implies [Fe/H] = −1.5. Hence, we may infer that the bulk of the Sculptor RR Lyr stars have metallicities equal or lower than $-1.5$. The Butler-Blanco scale is 0.2 dex more metal rich than the cluster scale of Zinn and West (1984). Hence, distribution of the Sculptor RRab stars shows cut-off at the period corresponding to [Fe/H] $\approx -1.7$ on Zinn-West metallicity scale. Our sample of RRab stars includes two variables with very short periods, V3761 with $P = 0.384$ days and V5330 with $P = 0.392$ days. Based on Eq. 3 we arrive to conclusion that metallicity of these stars is [Fe/H] $\leq -0.7$ on the scale of Zinn and West. The light curve of V5330 exhibits Blazhko effect.

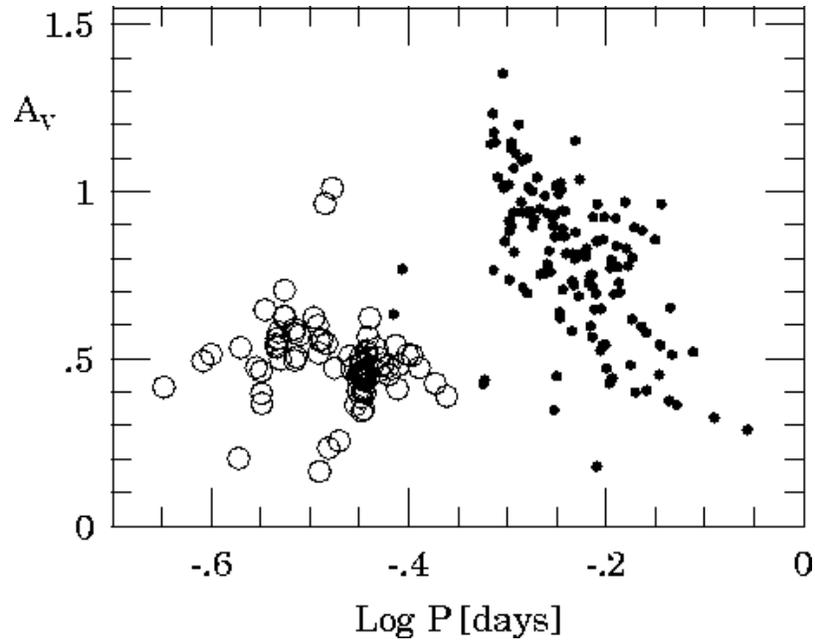

Figure 3: Period − V amplitude diagram for RR Lyr stars in the Sculptor dwarf galaxy. Filled circles − RRab variables, open circles − RRc variables.

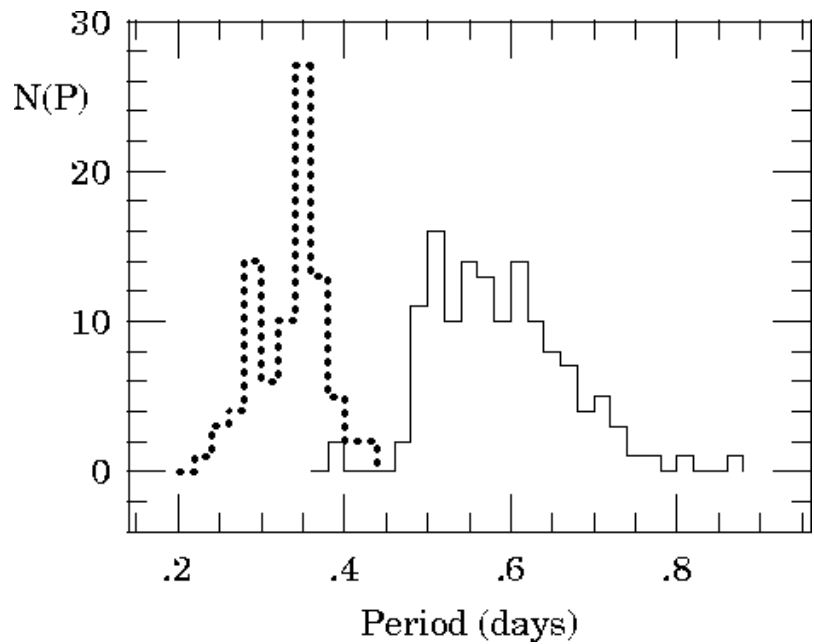

Figure 4: Period distribution of the RR Lyr stars in the Sculptor dwarf galaxy. Distribution for RRab stars is marked with the solid line while distribution for RRc stars is marked with the dashed line. Bins are 0.025 days wide.

In Fig. 5 we show a period vs. average V magnitude diagram. Three variables with periods close to 0.5 day whose light curves had poorly defined maxima (V2058, V2558 and V4785) were not taken into account. Average V magnitudes listed for these three stars in Table 1 are undoubtedly overestimated. Average V magnitudes of RRab stars show clear correlation with periods. In particular, the lower envelope of the $< V >$ distribution moves toward brighter magnitudes for longer periods. In their recent study Nemec et al. (1994) found a slope $\Delta M_{\mathrm{V}}/\Delta log P = -0.52 \pm 0.11$ for the $P - L - [Fe/H]$ relation in V. Using the data shown in Fig. 5 we obtained $\Delta M_{\mathrm{V}}/\Delta log P \approx -1.7$. Most likely the correlation between $< V >$ and $P$ observed for the Sculptor RRab variables is in large part due to metallicity spread exhibited by stars in this dwarf galaxy (Kunkel and Demers 1977, Da Costa 1984). Absolute magnitudes of RRab variables are strongly correlated with metallicity, $M_{\mathrm{V}}$ decreases with decreasing metallicity. On the other hand, average periods of RRab stars are longer for lower metallicities. Hence, for a mixture of RRab variables with different metallicities we may expect presence of stronger correlation between $< M_{\mathrm{V}} >$ and a period than it is observed for samples of variables with constant metallicity.

The average values of $< V >$ are $20.14 \pm 0.09$ and $20.12 \pm 0.11$ for RRab and RRc variables respectively. Adopting [Fe/H] $= -1.7$ ([Fe/H] $= -1.9$ one the Zinn-West scale) for the average metallicity of Sculptor RRab stars (see Section 4) and using relation $< M_{\mathrm{V}} >= 0.30 \times$ [Fe/H] $+ 0.94$ derived by Sandage (1993b) we obtain $(m - M)_{\mathrm{V}} = 19.71$ for the apparent distance modulus of the galaxy. This value is slightly larger than $(m - M)_{\mathrm{V}} = 19.56$ obtained by Kunkel and Demers (1977).

### 3.2. Anomalous cepheids and other bright variables

Anomalous cepheids are pulsating variables which are 1-2 mag brighter than RR Lyr stars. Their periods range from about 9 hours to about 38 hours. Anomalous cepheids were discovered in several dwarf spheroidal systems and in the Small Magellanic Cloud. Only one is known in galactic globular clusters. Nemec et al. (1994) published recently an extensive summary of data for known anomalous cepheids. Masses of anomalous cepheids are estimated at 1-2 solar masses. It is believed that these stars are created by mass transfer in binary systems. Three anomalous cepheids were discovered in the Sculptor galaxy so far (see van Agt 1978). Two of these stars were originally labeled A and B by Baade and Hubble (1939). The third one was discovered by Thackeray (1950). Smith and Stryker (1986) obtained low-resolution spectra of the Sculptor anomalous cepheids and determined their metallicities. They obtained [Fe/H] $= -1.9 \pm 0.2$, $-1.8 \pm 0.2$ and $-2.2 \pm 0.3$ for V25, V26 and V119, respectively (here variables are identified by van Agt's numbers).

The field monitored by us contained two previously known anomalous cepheids, namely variables numbered V26 and V119 by van Agt (1978). One new variable of this type (V5689) was discovered in our data. This new variable forms a close visual binary with a brighter star located at a distance of just 2.1 arcsec. This precluded probably its discovery during previous surveys of the Sculptor galaxy. The light curve parameters derived for anomalous cepheids observed by us are given in Table 3. This table contains the same quantities as Table 1. Periods derived for V734 and V3302 agree with periods listed by van Agt (1978) and by Swope (1968). The light curves of three observed anomalous cepheids are shown in Fig. 6. Photometry of variable V5689 is relatively noisy due to presence of bright visual companion to this star. Table 3 contains also data for two low-amplitude variables V274 and V687. These stars showed changes of brightness not exceeding 0.20 mag and with characteristic time scale of about 30 days. Our data are too sparse to decide if the observed variability is periodic. Both variables are located at the tip of the red giant branch of the galaxy (see next section). Their light curves for 1993 season are shown in Fig. 7.

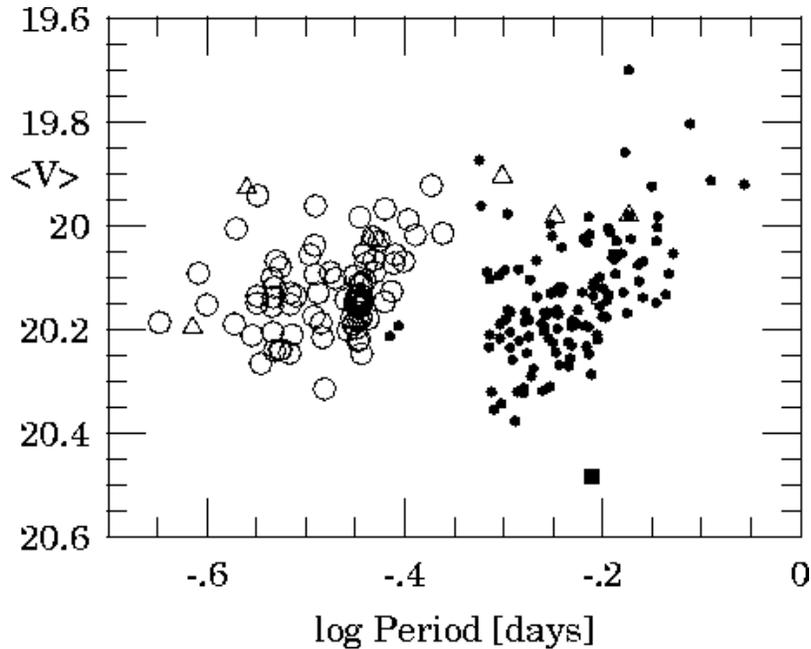

Figure 5: Average V magnitude vs. period for Sculptor RR Lyr stars. RRab stars are marked with filled circles while RRc stars are marked with open circles. Variables with Blashko effect are marked with open triangles. Position of deviating variable V4686 is shown with a filled square.

The equatorial coordinates of variables discussed in this section can be found at the end of Table 2.

## 4.  The color-magnitude diagram

The first color-magnitude diagram (hereafter CMD) of Sculptor reaching the horizontal branch was published by Kunkel and Demers (1977). They obtained photographic photometry for the field covering central part of the galaxy. Their CMD showed a predominantly red horizontal and wide red giant branch indicating presence of a significant abundance spread among Sculptor stars. Norris and Bessell (1978) suggested, based on the data of Kunkel and Demers (1977), that the abundance range present among Sculptor giants was equal to the abundance difference between the globular clusters M92 and M3, i.e. approximately 0.6 dex. Subsequently Da Costa (1984) obtained deep CMD reaching to the main-sequence turnoff for the relatively small field located just out of the core radius of Sculptor. His data confirmed finite width of the red giant branch of the galaxy. Based on comparison with theoretical isochrones Da Costa (1984) suggested that the bulk of the Sculptor stars may be 2-3 Gyr younger than stars in galactic globular clusters.

In Fig. 8 we show V vs. V-I and I vs. V-I CMDs of Sculptor. Stars which are known RR Lyr variables were not plotted in these figures. We marked however, positions of three anomalous cepheids and two bright variables discussed in Section 3.2.   In contrast to the photometry of Kunkel and Demers (1977) our CMD of Sculptor does show a relatively well populated blue horizontal branch (HB). The limiting magnitude of the I photometry coincides with the observed faint-blue end of the HB. Therefore, we cannot rule out the possibility that the blue part of the HB of Sculptor extends in fact toward fainter magnitudes than it is implied by data shown in Fig. 8. Morphology of the HB can be described with the index $(B-R)/(B+V+R)$ which was introduced by Lee (1989). $B$, $V$ and $R$ are the numbers of blue, variable and red HB stars, respectively. Using data presented in Fig. 8 we obtained $B = 288$, $V = 226$ and $R = 431$. The following limits on the location of the red HB stars were assumed: $19.95 < V < 20.26$ and $0.50 < V - I < 0.81$. The red part of the HB merges in fact with the RGB and therefore it is likely that we overestimated slightly the value of $R$. At the same time the value of $B$ is probably underestimated. Hence, our determination of $(B-R)/(B+V+R) = -0.22$ for Sculptor should be regarded as a lower limit on its HB index. However, even if we missed as many as 50% stars from the blue HB (what we consider very unlikely based on the unpublished photometry of Sculptor obtained recently by Kaluzny and Krzemiński) the HB index would rise only to $(B-R)/(B+V+R) = +0.1$. Location of Sculptor on the [Fe/H] vs. HB-type diagram indicates that its age is similar to ages of the relatively young globular clusters from the outer galactic halo (see Fig. 7 in Lee et al. 1994). This conclusion is in agreement with Da Costa (1984) who found, based on photometry of turn-off stars, that the bulk of Sculptor stars are younger than stars in M92.

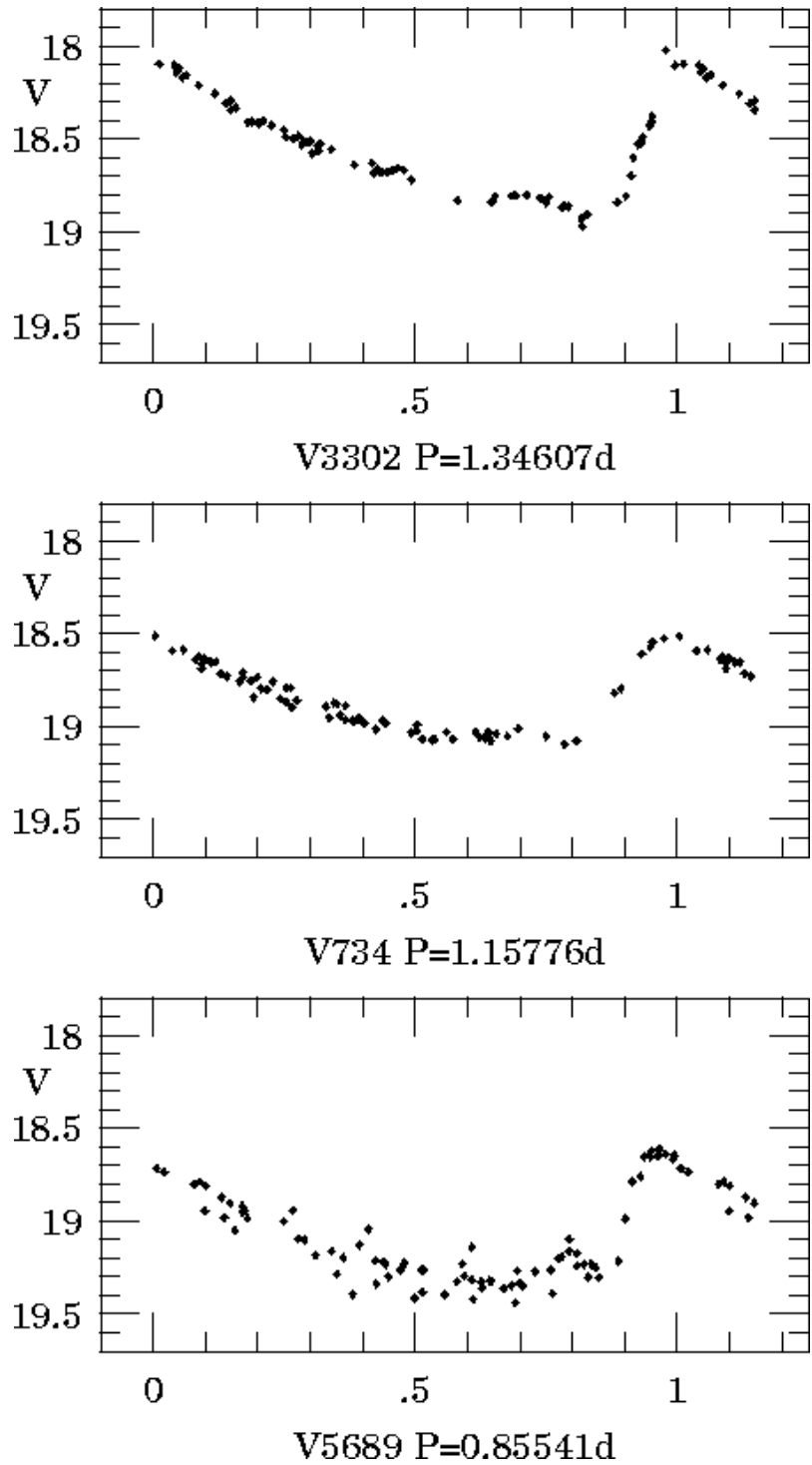

Figure 6: Phased light curves of Sculptor anomalous cepheids. Note change of the average magnitude with a period.

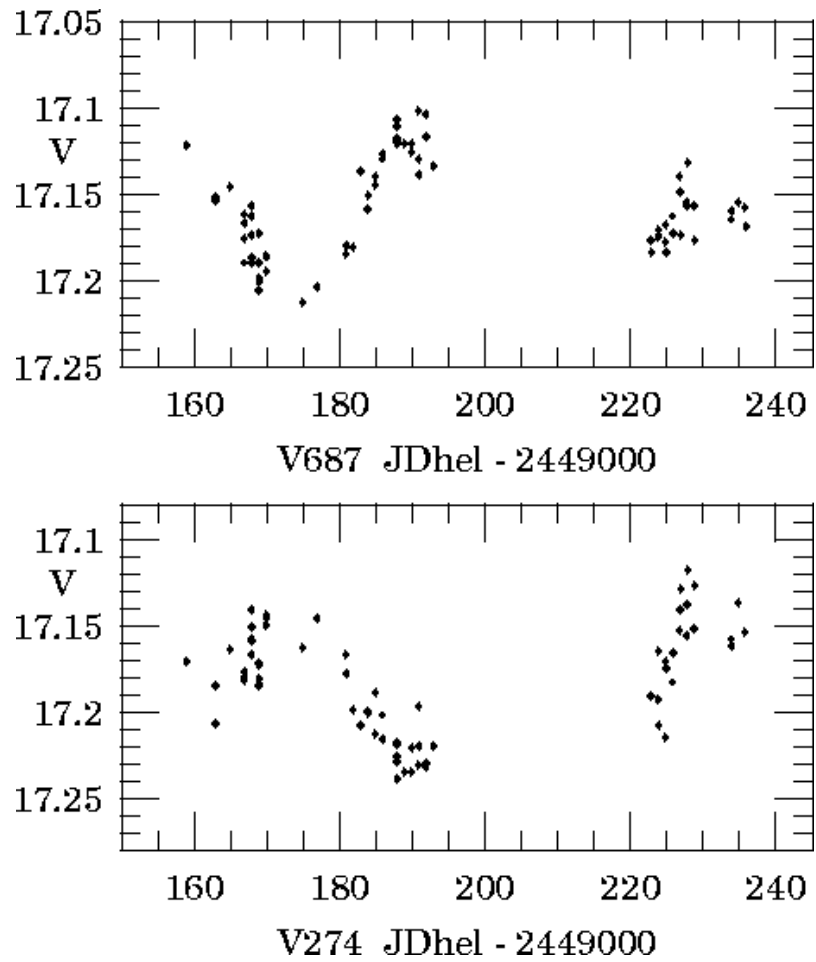

Figure 7: Light curves of two variable stars from the tip of the giant branch of Sculptor

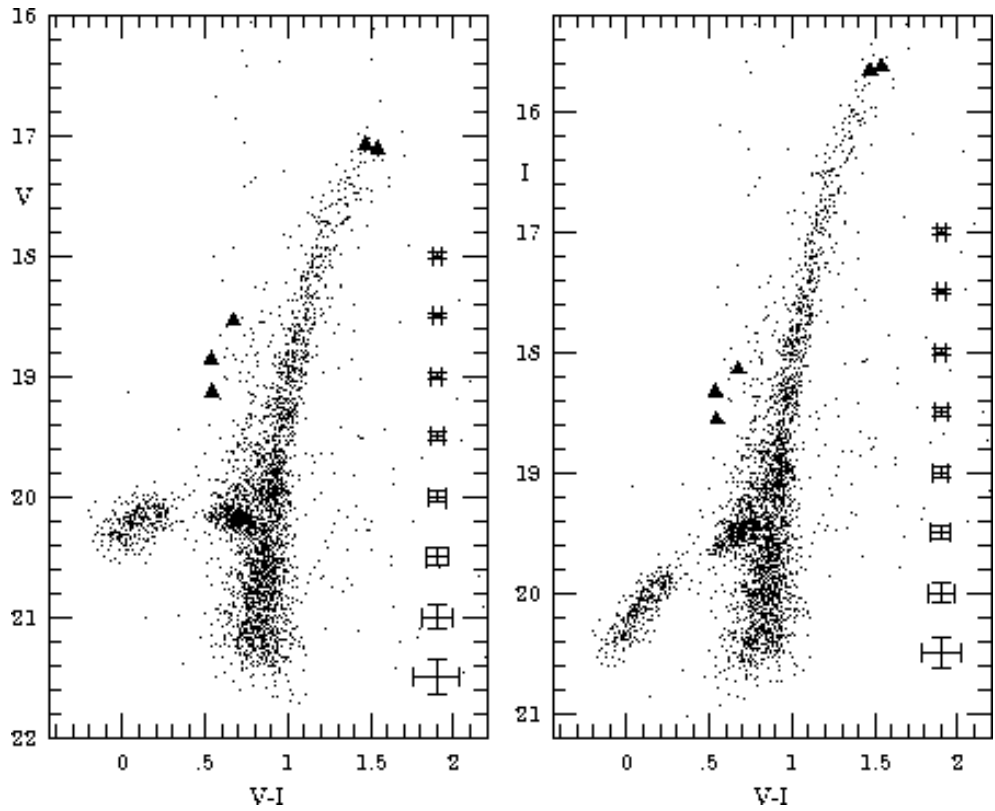

Figure 8: The V/V-I and I/V-I Color-magnitude diagrammes for the central part of the Sculptor dwarf galaxy. Stars known to be RR Lyr variables were not plotted. Positions of three anomalous cepheids and two bright variables of unknown type are marked with triangles. The adopted V magnitudes of variables correspond to their average V magnitudes (Table 3). The V-I colors and I magnitudes of variables were measured at some random phases of their light curves. The error bars shown indicate the size ($\pm 1\sigma$) of the formal errors for stars from the red giant branch.

In Fig. 9 we show CMD of Sculptor with over-imposed fiducials for giant branches of globular clusters M2 and M15 (Da Costa and Armandroff 1990). The fiducials were shifted assuming $E(V - I) = 0.03$ and $(m - M)_V = 19.71$. The galactic coordinates of Sculptor are $(l,b) = (283, -83)$ and therefore we adopted for it a reddening appropriate for the areas near the South galactic pole. On the Zinn-West scale metallicities of M2 and M15 are [Fe/H] $= -1.58$ and [Fe/H] $= -2.17$, respectively. The upper part of the red giant branch of Sculptor is bracketed by relations for M2 and M15. This indicates, that most of Sculptor stars have metallicities $-2.2 < $ [Fe/H] $ < -1.6$. This result is consistent with earlier estimates (Da Costa 1984, Norris and Bessell 1978) and with average value of metallicity obtained for RRab stars in Section 3 of this paper. The boundaries of the lower part of the RGB of Sculptor are shifted slightly to the blue relatively to relations defined by M2 and M15. This effect can be explained by a slightly lower age of bulk of Sculptor stars in comparison with ages of M2 and M15. Concluding this section we note, that significant fraction of stars which are located to the blue of the RGB in Fig. 9 and have $18.3 < V < 20.0$ are probably stars from the asymptotic giant branch of Sculptor.

## 5. Summary

The global properties of the Sculptor galaxy inferred from our data are in agreement with results of the earlier studies. Distribution of periods for RRab stars gives [Fe/H] $\leq -1.7$ for the bulk of stars from the horizontal branch. The range of colors exhibited by stars from the upper part of the red giant branch implicates spread in [Fe/H] from about -2.2 to about -1.6. Very similar results were obtained based on the analysis of the CMD of Sculptor by Da Costa (1984) and by Norris and Bessell (1978). Also metallicities obtained for 3 Sculptor anomalous cepheids by Smith and Stryker (1986) are within the quoted range of [Fe/H]. Morphology of the horizontal branch of Sculptor connected with the relatively low average metallicity of its stars indicates that age of the galaxy is close to the ages of youngest galactic globulars. This agrees with conclusion of Da Costa (1984) who found that turnoff of Sculptor is redder than turnoff of M92 although both objects have similar metallicities.

Besides Sculptor some detailed data about population of RR Lyr variables are available for three other dwarf galaxies from the Local Group: Ursa Minor (Nemec et al. 1988), Draco (Nemec 1985) and Carina (Saha et al. 1986). These data are summarized in Table 4 where we listed also metallicities adopted for each galaxy. According to Stetson (1984) stars in Ursa Minor show small range of metallicity and age. Carina contains two populations of stars with ages $\tau > 10$ and $\tau \approx 6$ Gyr. Metallicity of the older population is estimated at [Fe/H] = $-2.2$ (Smecker–Hane et al. 1994). Only older population of Carina may form RR Lyr variables. The Draco stars seem to form two groups in respect of metallicity, one with an average [Ca, Mg/H] near $-1.6 \pm 0.2$, and the other $-2.3 \pm 0.2$ (Lehnert et al. 1992). Six out of fourteen giants observed by Lehnert et al. (1992) belonged to the low-abundance group. Therefore, we may expect that significant fraction of stars from the HB of Draco have [Fe/H] $\approx -2.2$. The CMD of Draco published by Carney & Seitzer (1986) does not show any extended blue horizontal branch typical for very old globulars. Most probably Draco RR Lyr variables belong to the group of low-abundance stars. Figure 10 shows distribution of periods for RRab and RRc stars in Carina, Draco and Ursa Minor galaxies (distribution for Sculptor is shown in Fig. 4). We plotted in Fig. 10 also distributions of periods for variables from globular clusters M68 (Walker 1994) and Ru 106 (Kaluzny et al. 1995). M68 belongs to the group of the oldest clusters while Ru 106 is one of the youngest galactic globular clusters known (eg. Lee et al. 1994). Metallicities of M68 and Ru 106 are estimated at $-2.1$ and $-1.9$, respectively. Those values are similar to average metallicities of RR Lyr populations in four discussed dwarf galaxies. Data presented in Figs. 4 and 10 and in Table 4 indicate that ages of RR Lyr populations in Sculptor and Ursa Minor are comparable. Ursa Minor has larger value of $< P_{\mathrm{ab}} >$ and longer value of $P_{\mathrm{ab}}$ at the blue edge of the instability strip in comparison with Sculptor. This suggests that RR Lyr variables in Ursa Minor have lower average metallicity than Sculptor variables. In all four galaxies distributions of periods of RRc variables show maxima near $P = 0.40$ days. It is interesting that Sculptor and Ursa Minor show secondary peaks at $P \approx 0.32$ days. Such bimodal distributions are not common among globular clusters. Data presented in Table 4 indicate that populations of RR Lyr stars in Sculptor and Ursa Minor are older than Ru 106 but younger than M68. RR Lyr stars in Carina and Draco seem to be coeval or only marginally older than Ru 106.

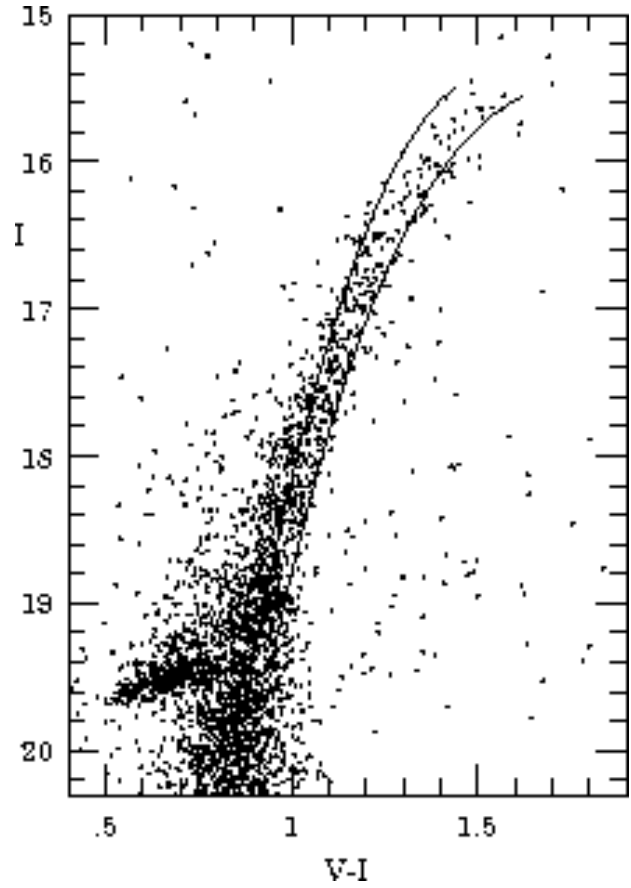

Figure 9: The I/V-I color-magnitude diagram of Sculptor. Stars known to be variables were not plotted. The solid lines show location of red giant branches for M2 and M15.

## Acknowledgements

This project was supported by NSF grants AST 92-16494 to Bohdan Paczynski and AST 92-16830 to George Preston, and by the Polish KBN grant BST 475/A/94 to Marcin Kubiak. JK was supported also by KBN grant PB30400506.

## Appendix A

Tables containing light curves of all variables discussed in this paper are published by A&A at the Centre de Donneess de Strasbourg, where they are available in electronic form: See the editorial in A&A 1993, Vol. 280, page E1. We also submitted to the data base an image allowing identification of all variables.

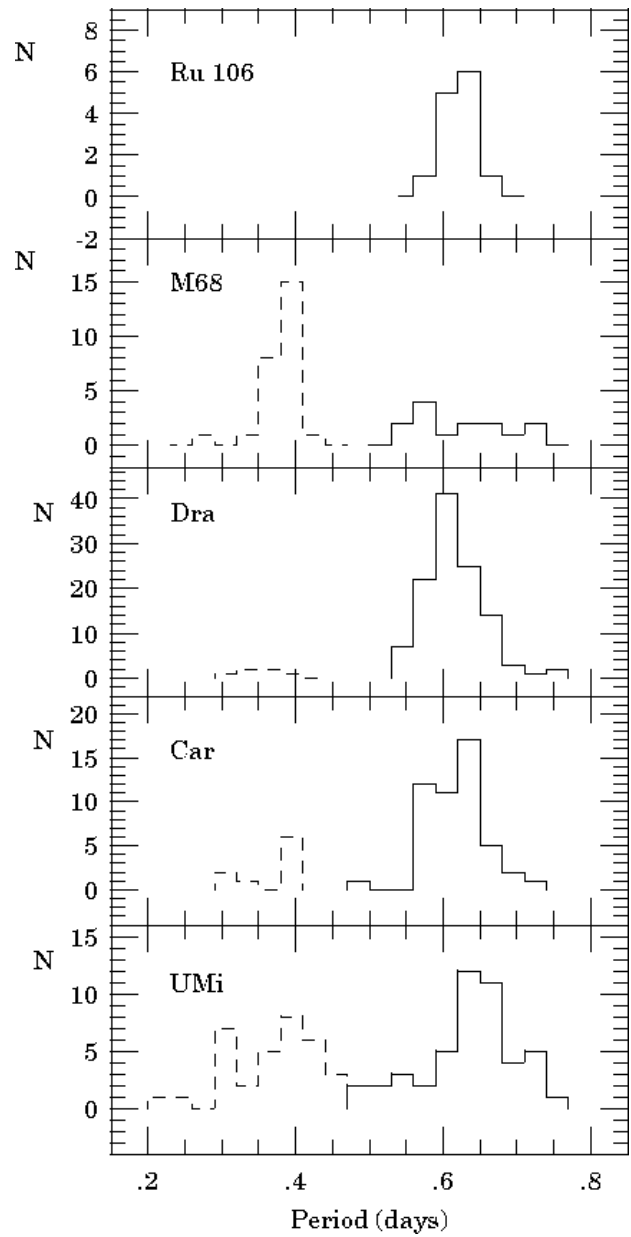

Figure 10: Period distribution of the RR Lyr stars in Carina, Draco and Ursa Minor spheroidal galaxies. Distributions for globular clusters M68 and Ru 106 are also shown. Bins are 0.03 days wide. Solid line corresponds to RRab variables while dashed line corresponds to RRc variables. Variables of type RRd were omitted.

Table 1: Light curve parameters and rectangular coordinates of Sculptor RR Lyr variables. Coordinates X and Y are expressed in pixels (1 pixel equals 0.435"). See text for more details.

| ID | P(days) | Nv | Vmax | Vmin | Av | < V > | X | Y | type |
|----|---------|-----|-------|-------|------|--------|-------|--------|------|
| 33 | 0.30906 | 80 | 19.86 | 20.43 | 0.57 | 20.137 | 141.0 | 27.2 | c |
| 37 | 0.50871 | 82 | 19.65 | 20.58 | 0.94 | 20.237 | 205.2 | 83.6 | ab |
| 38 | 0.37123 | 82 | 19.90 | 20.36 | 0.46 | 20.091 | 425.2 | 108.2 | c |
| 54 | 0.61450 | 82 | 19.89 | 20.54 | 0.65 | 20.288 | 272.1 | 321.2 | ab |
| 59 | 0.35969 | 82 | 19.95 | 20.39 | 0.44 | 20.157 | 435.4 | 357.4 | c |
| 71 | 0.43494 | 82 | 19.85 | 20.24 | 0.39 | 20.016 | 125.3 | 405.9 | c |
| 116 | 0.73599 | 82 | 19.79 | 20.30 | 0.51 | 20.094 | 493.9 | 281.2 | ab |
| 321 | 0.60184 | 82 | 19.70 | 20.50 | 0.80 | 20.130 | 98.0 | 610.6 | ab |
| 356 | 0.56950 | 83 | 19.57 | 20.51 | 0.94 | 20.120 | 73.0 | 810.7 | ab |
| 357 | 0.55808 | 83 | 19.72 | 20.61 | 0.90 | 20.312 | 223.4 | 814.2 | ab |
| 361 | 0.29759 | 83 | 19.92 | 20.55 | 0.63 | 20.241 | 111.4 | 850.4 | c |
| 366 | 0.68510 | 83 | 19.72 | 20.31 | 0.60 | 20.077 | 296.2 | 874.6 | ab |
| 368 | 0.35896 | 83 | 19.91 | 20.37 | 0.46 | 20.154 | 233.3 | 875.8 | c |
| 377 | 0.29522 | 83 | 19.84 | 20.33 | 0.50 | 20.069 | 491.4 | 927.0 | c |
| 385 | 0.62854 | 83 | 19.50 | 20.42 | 0.92 | 20.141 | 86.8 | 968.0 | abB |
| 403 | 0.33033 | 71 | 20.18 | 20.41 | 0.23 | 20.316 | 28.6 | 568.7 | c |
| 406 | 0.55025 | 83 | 19.77 | 20.56 | 0.78 | 20.195 | 499.1 | 580.1 | ab |
| 411 | 0.22499 | 83 | 19.95 | 20.37 | 0.41 | 20.188 | 245.5 | 594.2 | c |
| 416 | 0.52602 | 83 | 19.56 | 20.58 | 1.01 | 20.187 | 206.6 | 603.2 | ab |
| 439 | 0.49612 | 78 | 19.34 | 20.69 | 1.35 | 20.097 | 290.9 | 764.6 | ab |
| 462 | 0.52494 | 83 | 19.52 | 20.62 | 1.10 | 20.324 | 79.9 | 848.1 | ab |
| 463 | 0.56842 | 83 | 19.55 | 20.55 | 1.00 | 20.200 | 180.7 | 847.8 | ab |
| 493 | 0.68687 | 83 | 19.58 | 20.46 | 0.88 | 20.109 | 71.3 | 974.9 | ab |
| 737 | 0.33863 | 82 | 19.94 | 20.20 | 0.25 | 20.103 | 102.4 | 1005.5 | c |
| 753 | 0.49002 | 83 | 19.51 | 20.56 | 1.04 | 20.356 | 308.2 | 1084.6 | ab |
| 763 | 0.29312 | 83 | 19.89 | 20.43 | 0.53 | 20.139 | 265.9 | 1146.5 | c |
| 771 | 0.81187 | 83 | 19.74 | 20.06 | 0.32 | 19.914 | 75.3 | 1177.7 | ab |
| 782 | 0.48550 | 83 | 19.45 | 20.63 | 1.18 | 20.104 | 111.0 | 1222.5 | ab |
| 786 | 0.27909 | 83 | 19.96 | 20.43 | 0.47 | 20.213 | 277.7 | 1252.3 | c |
| 793 | 0.35993 | 83 | 19.92 | 20.37 | 0.44 | 20.136 | 266.7 | 1304.1 | c |
| 803 | 0.57363 | 83 | 19.49 | 20.43 | 0.94 | 20.043 | 369.0 | 1370.2 | ab |
| 811 | 0.36314 | 83 | 19.92 | 20.37 | 0.45 | 20.107 | 442.3 | 1413.9 | c |
| 853 | 0.51955 | 71 | 19.75 | 20.47 | 0.71 | 20.223 | 59.3 | 1116.8 | ab |
| 860 | 0.56222 | 83 | 20.00 | 20.45 | 0.45 | 20.227 | 109.5 | 1160.1 | ab |
| 1132 | 0.63158 | 79 | 19.88 | 20.35 | 0.47 | 20.122 | 352.0 | 1534.8 | ab |
| 1139 | 0.51892 | 79 | 19.37 | 20.47 | 1.09 | 20.086 | 449.8 | 1616.1 | ab |
| 1142 | 0.67062 | 79 | 19.39 | 20.00 | 0.62 | 19.701 | 364.0 | 1637.9 | ab |
| 1149 | 0.32342 | 79 | 19.86 | 20.03 | 0.16 | 19.963 | 303.4 | 1690.8 | c |
| 1164 | 0.60621 | 79 | 19.80 | 20.53 | 0.73 | 20.194 | 177.3 | 1819.3 | ab |
| 1168 | 0.40358 | 79 | 19.74 | 20.21 | 0.48 | 19.982 | 274.7 | 1891.5 | d |
| 1178 | 0.42362 | 73 | 19.74 | 20.16 | 0.43 | 19.924 | 77.2 | 2021.5 | c |
| 1204 | 0.29233 | 79 | 19.83 | 20.37 | 0.54 | 20.102 | 108.8 | 1616.3 | c |
| 1256 | 0.58717 | 79 | 19.69 | 20.49 | 0.80 | 20.233 | 115.4 | 1966.0 | ab |
| 1261 | 0.63037 | 78 | 19.86 | 20.40 | 0.54 | 20.177 | 450.0 | 2001.0 | ab |
| 1411 | 0.66458 | 80 | 19.46 | 20.24 | 0.78 | 19.860 | 630.6 | 74.8 | ab |
| 1424 | 0.61230 | 79 | 19.44 | 20.36 | 0.92 | 20.018 | 963.6 | 159.5 | ab |
| 1439 | 0.35606 | 75 | 19.93 | 20.39 | 0.46 | 20.145 | 1000.0 | 270.3 | c |
| 1446 | 0.77331 | 80 | 19.53 | 20.05 | 0.52 | 19.805 | 908.3 | 311.7 | ab |
| 1457 | 0.71782 | 80 | 19.40 | 20.36 | 0.96 | 19.983 | 717.5 | 375.3 | ab |
| 1462 | 0.56113 | 80 | 19.40 | 20.41 | 1.02 | 20.021 | 846.3 | 410.4 | ab |
| 1470 | 0.50565 | 80 | 19.69 | 20.59 | 0.90 | 19.979 | 846.8 | 451.5 | ab |
| 1482 | 0.29810 | 79 | 19.71 | 20.42 | 0.70 | 20.081 | 954.0 | 466.4 | c |
| 1491 | 0.35786 | 80 | 19.96 | 20.36 | 0.40 | 20.154 | 541.6 | 77.3 | c |
| 1519 | 0.35685 | 80 | 19.95 | 20.43 | 0.48 | 20.178 | 719.8 | 239.2 | c |
| 1546 | 0.53124 | 80 | 19.52 | 20.52 | 1.00 | 20.215 | 678.1 | 363.6 | ab |
| 1553 | 0.71644 | 80 | 19.69 | 20.23 | 0.54 | 20.004 | 986.1 | 408.0 | ab |
| 1555 | 0.52724 | 80 | 19.59 | 20.53 | 0.93 | 20.168 | 625.2 | 411.2 | ab |
| 1558 | 0.24301 | 80 | 19.92 | 20.41 | 0.49 | 20.198 | 513.4 | 424.3 | cB |
| 1566 | 0.57028 | 80 | 19.56 | 20.45 | 0.89 | 20.172 | 637.6 | 448.2 | ab |
| 1823 | 0.29846 | 83 | 19.94 | 20.56 | 0.63 | 20.240 | 541.6 | 774.6 | c |
| 1824 | 0.47550 | 83 | 19.70 | 20.14 | 0.44 | 19.963 | 595.6 | 782.9 | ab |
| 1830 | 0.51784 | 83 | 19.64 | 20.61 | 0.97 | 20.322 | 852.2 | 792.3 | ab |
| 1838 | 0.70752 | 83 | 19.51 | 20.36 | 0.86 | 19.926 | 742.5 | 841.1 | ab |
| 1873 | 0.29230 | 83 | 19.88 | 20.42 | 0.54 | 20.157 | 621.2 | 956.7 | c |
| 1874 | 0.26901 | 83 | 19.76 | 20.30 | 0.53 | 20.007 | 686.7 | 956.5 | c |
| 1875 | 0.33319 | 83 | 19.48 | 20.56 | 1.01 | 20.039 | 923.3 | 956.6 | cB |

Table 1: - continued

| ID | P(days) | Nv | Vmax | Vmin | Av | < V > | X | Y | type |
|----|---------|-----|------|------|-----|-------|-----|-----|------|
| 1877 | 0.56716 | 82 | 19.79 | 20.41 | 0.62 | 20.171 | 588.9 | 964.2 | ab |
| 1890 | 0.36376 | 83 | 19.86 | 20.48 | 0.62 | 20.074 | 744.8 | 541.8 | c |
| 1899 | 0.64671 | 83 | 19.49 | 20.33 | 0.84 | 20.056 | 567.8 | 519.6 | ab |
| 1910 | 0.57285 | 83 | 19.59 | 20.46 | 0.87 | 20.130 | 854.5 | 539.7 | ab |
| 1914 | 0.57051 | 80 | 19.80 | 20.50 | 0.71 | 20.270 | 805.0 | 548.2 | ab |
| 1926 | 0.55025 | 83 | 19.79 | 20.55 | 0.77 | 20.204 | 499.1 | 580.1 | ab |
| 1930 | 0.61118 | 83 | 19.91 | 20.47 | 0.56 | 20.249 | 886.7 | 605.0 | ab |
| 1932 | 0.50604 | 83 | 19.42 | 20.56 | 1.15 | 20.164 | 553.3 | 607.1 | ab |
| 1940 | 0.69306 | 83 | 19.81 | 20.39 | 0.58 | 20.140 | 588.5 | 633.6 | ab |
| 1941 | 0.36567 | 83 | 19.92 | 20.40 | 0.47 | 20.181 | 719.4 | 633.2 | c |
| 1943 | 0.55116 | 83 | 19.61 | 20.54 | 0.93 | 20.169 | 976.4 | 643.5 | ab |
| 1997 | 0.62674 | 83 | 19.64 | 20.49 | 0.86 | 20.101 | 904.4 | 823.9 | ab |
| 2004 | 0.58735 | 83 | 19.38 | 20.54 | 1.15 | 20.196 | 882.0 | 842.2 | ab |
| 2012 | 0.71475 | 83 | 19.89 | 20.35 | 0.45 | 20.150 | 905.2 | 855.9 | ab |
| 2021 | 0.62292 | 83 | 19.86 | 20.39 | 0.52 | 20.212 | 784.0 | 876.0 | ab |
| 2048 | 0.35836 | 83 | 19.84 | 20.44 | 0.61 | 20.150 | 797.4 | 979.6 | cB |
| 2058 | 0.50350 | 83 | 19.72 | 20.60 | 0.88 | 20.404 | 973.2 | 999.8 | ab |
| 2059 | 0.49692 | 83 | 19.49 | 20.51 | 1.02 | 20.191 | 729.6 | 1001.1 | ab |
| 2410 | 0.53183 | 84 | 19.67 | 20.57 | 0.89 | 20.183 | 788.3 | 1014.8 | ab |
| 2421 | 0.40831 | 84 | 19.80 | 20.27 | 0.47 | 20.019 | 678.1 | 1060.6 | c |
| 2422 | 0.48446 | 84 | 19.49 | 20.72 | 1.23 | 20.236 | 957.5 | 1061.0 | ab |
| 2423 | 0.55906 | 84 | 19.85 | 20.20 | 0.35 | 19.998 | 998.7 | 1063.6 | ab |
| 2424 | 0.34878 | 84 | 19.93 | 20.40 | 0.47 | 20.203 | 546.7 | 1067.1 | c |
| 2425 | 0.87792 | 84 | 19.81 | 20.10 | 0.29 | 19.922 | 647.2 | 1072.4 | ab |
| 2450 | 0.61802 | 84 | 19.54 | 20.50 | 0.96 | 20.112 | 671.9 | 1148.4 | ab |
| 2455 | 0.63622 | 82 | 19.92 | 20.35 | 0.43 | 20.178 | 512.6 | 1181.8 | ab |
| 2458 | 0.35769 | 84 | 19.94 | 20.39 | 0.45 | 20.185 | 828.3 | 1191.6 | c |
| 2467 | 0.35809 | 84 | 19.93 | 20.34 | 0.41 | 20.133 | 969.7 | 1222.5 | c |
| 2470 | 0.69339 | 84 | 19.85 | 20.26 | 0.41 | 20.070 | 743.2 | 1230.3 | ab |
| 2471 | 0.65023 | 84 | 19.71 | 20.44 | 0.73 | 20.094 | 820.6 | 1233.7 | ab |
| 2482 | 0.26786 | 84 | 20.11 | 20.31 | 0.20 | 20.191 | 543.8 | 1251.6 | c |
| 2502 | 0.32769 | 82 | 19.69 | 20.65 | 0.96 | 20.189 | 523.7 | 1335.5 | c |
| 2528 | 0.60323 | 84 | 19.50 | 20.33 | 0.83 | 20.027 | 869.6 | 1392.1 | ab |
| 2545 | 0.67408 | 84 | 19.50 | 20.39 | 0.89 | 20.027 | 821.1 | 1478.2 | ab |
| 2552 | 0.53444 | 84 | 19.64 | 20.55 | 0.92 | 20.292 | 656.1 | 1096.3 | ab |
| 2555 | 0.50272 | 84 | 19.50 | 20.52 | 1.02 | 20.087 | 660.8 | 1261.9 | ab |
| 2558 | 0.50340 | 84 | 19.71 | 20.62 | 0.91 | 20.419 | 973.2 | 999.8 | ab |
| 2559 | 0.49692 | 84 | 19.51 | 20.52 | 1.01 | 20.219 | 729.5 | 1001.0 | ab |
| 2562 | 0.38627 | 83 | 19.88 | 20.42 | 0.54 | 20.128 | 907.8 | 1006.0 | c |
| 2566 | 0.58356 | 84 | 19.77 | 20.58 | 0.81 | 20.267 | 592.7 | 1011.0 | ab |
| 2575 | 0.61108 | 84 | 19.52 | 20.27 | 0.75 | 19.984 | 578.0 | 1035.2 | ab |
| 2606 | 0.58531 | 84 | 19.81 | 20.53 | 0.72 | 20.257 | 719.1 | 1132.1 | ab |
| 2627 | 0.57508 | 84 | 19.64 | 20.46 | 0.81 | 20.121 | 529.4 | 1209.0 | ab |
| 2639 | 0.28961 | 84 | 19.93 | 20.47 | 0.54 | 20.212 | 739.1 | 1254.4 | c? |
| 2689 | 0.51136 | 84 | 19.46 | 20.57 | 1.11 | 20.260 | 740.2 | 1480.7 | ab |
| 2699 | 0.48517 | 84 | 19.73 | 20.50 | 0.76 | 20.212 | 668.1 | 1497.1 | ab |
| 2991 | 0.55264 | 81 | 19.63 | 20.45 | 0.82 | 20.175 | 747.1 | 1526.2 | ab |
| 3004 | 0.71550 | 81 | 19.75 | 20.30 | 0.54 | 20.031 | 653.2 | 1599.7 | ab |
| 3009 | 0.36013 | 81 | 19.72 | 20.25 | 0.52 | 19.971 | 890.2 | 1625.4 | ab? |
| 3016 | 0.36031 | 81 | 19.90 | 20.35 | 0.46 | 20.111 | 811.8 | 1638.8 | c |
| 3019 | 0.73295 | 81 | 19.63 | 20.28 | 0.65 | 19.960 | 711.6 | 1650.5 | abB |
| 3024 | 0.36657 | 80 | 19.76 | 20.52 | 0.76 | 20.027 | 533.7 | 1699.2 | ?B |
| 3026 | 0.64546 | 81 | 19.50 | 20.42 | 0.92 | 20.063 | 881.8 | 1707.5 | ab |
| 3039 | 0.65198 | 81 | 19.59 | 20.29 | 0.70 | 20.031 | 973.3 | 1843.1 | ab |
| 3043 | 0.62390 | 81 | 19.83 | 20.48 | 0.65 | 20.221 | 983.6 | 1857.9 | ab |
| 3044 | 0.35393 | 81 | 19.90 | 20.31 | 0.40 | 20.098 | 617.3 | 1874.6 | c |
| 3104 | 0.35700 | 81 | 20.10 | 20.44 | 0.34 | 20.223 | 887.8 | 1763.8 | c |
| 3113 | 0.59324 | 81 | 19.51 | 20.55 | 1.03 | 20.191 | 724.3 | 1817.4 | ab |
| 3125 | 0.53310 | 81 | 19.53 | 20.45 | 0.92 | 20.106 | 991.1 | 1905.1 | ab |
| 3126 | 0.35294 | 81 | 19.97 | 20.42 | 0.44 | 20.191 | 946.6 | 1909.5 | c |
| 3143 | 0.35421 | 74 | 19.95 | 20.39 | 0.44 | 20.150 | 942.9 | 2014.4 | c |
| 3318 | 0.64023 | 75 | 19.90 | 20.34 | 0.44 | 20.134 | 1296.8 | 32.0 | ab |
| 3319 | 0.56498 | 75 | 19.48 | 20.47 | 0.99 | 19.984 | 1448.3 | 32.0 | abB |
| 3320 | 0.28247 | 75 | 19.94 | 20.33 | 0.40 | 20.137 | 1333.7 | 66.9 | c |
| 3345 | 0.35606 | 73 | 19.94 | 20.39 | 0.45 | 20.156 | 1000.0 | 270.3 | c |
| 3346 | 0.35753 | 75 | 19.92 | 20.37 | 0.45 | 20.115 | 1245.4 | 280.0 | c |
| 3365 | 0.66807 | 75 | 19.86 | 20.34 | 0.48 | 20.170 | 1034.2 | 347.6 | ab |

Table 1: - continued

| ID | P(days) | Nv | Vmax | Vmin | Av | < V > | X | Y | type |
|---|---|---|---|---|---|---|---|---|---|
| 3397 | 0.54092 | 75 | 19.48 | 20.43 | 0.95 | 20.068 | 1319.8 | 238.8 | ab |
| 3410 | 0.54146 | 75 | 19.66 | 20.41 | 0.75 | 20.139 | 1233.1 | 66.4 | ab |
| 3413 | 0.35955 | 75 | 19.93 | 20.36 | 0.43 | 20.115 | 1272.8 | 86.4 | c |
| 3468 | 0.29382 | 75 | 19.83 | 20.39 | 0.56 | 20.120 | 1118.1 | 428.8 | c |
| 3710 | 0.47382 | 84 | 19.67 | 20.10 | 0.42 | 19.875 | 1236.5 | 544.3 | ab |
| 3760 | 0.54851 | 84 | 19.75 | 20.50 | 0.76 | 20.320 | 1126.8 | 800.0 | ab |
| 3761 | 0.38447 | 84 | 19.80 | 20.44 | 0.63 | 20.214 | 1293.2 | 810.2 | ab |
| 3763 | 0.32214 | 84 | 19.90 | 20.50 | 0.60 | 20.176 | 1462.3 | 815.7 | c |
| 3777 | 0.63764 | 84 | 19.66 | 20.43 | 0.77 | 20.146 | 1237.7 | 881.2 | ab |
| 3801 | 0.36978 | 84 | 19.80 | 20.33 | 0.53 | 20.023 | 1285.6 | 969.5 | c |
| 3810 | 0.66197 | 84 | 19.61 | 20.43 | 0.83 | 20.130 | 1146.8 | 994.6 | ab |
| 3827 | 0.58776 | 84 | 19.62 | 20.50 | 0.88 | 20.185 | 1117.4 | 554.9 | ab |
| 3832 | 0.35651 | 83 | 19.96 | 20.34 | 0.38 | 20.158 | 1318.3 | 574.3 | c |
| 3834 | 0.37522 | 84 | 19.74 | 20.23 | 0.48 | 20.031 | 1431.4 | 580.3 | c |
| 3862 | 0.29408 | 84 | 19.95 | 20.53 | 0.58 | 20.242 | 1355.1 | 687.0 | c |
| 3888 | 0.60805 | 84 | 19.79 | 20.53 | 0.75 | 20.234 | 1350.5 | 791.0 | ab |
| 3907 | 0.58319 | 84 | 19.76 | 20.49 | 0.73 | 20.273 | 1124.2 | 856.2 | ab |
| 3916 | 0.30500 | 84 | 19.86 | 20.45 | 0.59 | 20.153 | 1026.9 | 887.9 | c |
| 3931 | 0.36016 | 83 | 19.94 | 20.46 | 0.52 | 20.248 | 1006.1 | 954.2 | c |
| 3934 | 0.51980 | 84 | 19.61 | 20.55 | 0.94 | 20.125 | 1141.8 | 975.3 | abB |
| 3938 | 0.37984 | 84 | 19.90 | 20.38 | 0.48 | 20.148 | 1326.5 | 989.1 | c |
| 3941 | 0.35695 | 84 | 19.91 | 20.37 | 0.46 | 20.148 | 1449.6 | 997.2 | c |
| 4233 | 0.35852 | 82 | 19.83 | 20.18 | 0.35 | 19.985 | 998.7 | 1063.6 | c |
| 4235 | 0.35566 | 83 | 19.92 | 20.42 | 0.50 | 20.199 | 1478.7 | 1080.4 | c |
| 4263 | 0.28478 | 73 | 19.91 | 20.55 | 0.65 | 20.267 | 1298.8 | 1219.2 | c |
| 4272 | 0.28239 | 84 | 19.95 | 20.41 | 0.46 | 20.151 | 1168.1 | 1238.9 | c |
| 4277 | 0.30630 | 84 | 19.91 | 20.42 | 0.50 | 20.212 | 1317.0 | 1268.8 | c |
| 4291 | 0.38799 | 84 | 19.88 | 20.28 | 0.41 | 20.074 | 1226.3 | 1324.7 | c |
| 4308 | 0.35896 | 84 | 19.92 | 20.38 | 0.46 | 20.135 | 1035.6 | 1391.2 | c |
| 4309 | 0.55430 | 82 | 19.68 | 20.44 | 0.76 | 20.169 | 1344.2 | 1394.4 | ab |
| 4313 | 0.73108 | 81 | 19.91 | 20.28 | 0.37 | 20.134 | 1421.8 | 1430.7 | ab |
| 4353 | 0.35981 | 84 | 20.01 | 20.41 | 0.40 | 20.178 | 1013.8 | 1104.1 | c |
| 4385 | 0.48740 | 84 | 19.46 | 20.61 | 1.15 | 20.322 | 1101.4 | 1243.5 | ab |
| 4686 | 0.61608 | 81 | 20.04 | 20.74 | 0.69 | 20.484 | 1427.1 | 1514.4 | ab |
| 4689 | 0.63920 | 81 | 19.57 | 20.36 | 0.80 | 20.006 | 1243.9 | 1527.7 | ab |
| 4747 | 0.59194 | 81 | 19.79 | 20.48 | 0.69 | 20.189 | 1409.1 | 1513.0 | ab |
| 4780 | 0.46386 | 81 | 19.49 | 20.33 | 0.84 | 19.912 | 1189.9 | 1660.9 | d |
| 4785 | 0.50611 | 80 | 19.48 | 20.60 | 1.13 | 20.266 | 1293.0 | 1701.9 | ab |
| 4786 | 0.53728 | 81 | 19.42 | 20.51 | 1.09 | 20.048 | 1393.8 | 1702.1 | ab |
| 4793 | 0.55984 | 81 | 19.50 | 20.43 | 0.93 | 20.133 | 1372.3 | 1756.2 | ab |
| 4812 | 0.48232 | 81 | 19.41 | 20.55 | 1.14 | 20.091 | 1343.7 | 1877.9 | ab |
| 4824 | 0.36224 | 81 | 19.73 | 20.30 | 0.56 | 20.054 | 1228.1 | 1988.7 | c |
| 5000 | 0.32325 | 71 | 19.88 | 20.43 | 0.56 | 20.095 | 1846.6 | 32.8 | c |
| 5011 | 0.27571 | 70 | 19.56 | 20.37 | 0.81 | 19.929 | 1543.5 | 147.6 | cB |
| 5015 | 0.37067 | 71 | 19.84 | 20.35 | 0.50 | 20.067 | 1665.6 | 188.1 | c |
| 5030 | 0.58271 | 71 | 19.91 | 20.49 | 0.58 | 20.226 | 1567.3 | 286.7 | ab |
| 5032 | 0.34702 | 71 | 19.90 | 20.40 | 0.51 | 20.144 | 1670.8 | 304.1 | c |
| 5049 | 0.64875 | 71 | 19.65 | 20.42 | 0.77 | 20.067 | 1848.8 | 391.6 | ab |
| 5065 | 0.38023 | 71 | 19.78 | 20.23 | 0.45 | 19.969 | 1699.6 | 460.2 | c |
| 5068 | 0.74402 | 71 | 19.88 | 20.25 | 0.36 | 20.055 | 1837.4 | 466.8 | ab |
| 5081 | 0.61752 | 71 | 20.02 | 20.19 | 0.18 | 20.120 | 1573.8 | 397.3 | ab |
| 5085 | 0.31931 | 51 | 19.82 | 20.44 | 0.62 | 20.057 | 1996.0 | 27.3 | c |
| 5105 | 0.55664 | 71 | 19.59 | 20.52 | 0.92 | 20.219 | 1784.8 | 157.9 | ab |
| 5123 | 0.32534 | 71 | 19.89 | 20.45 | 0.56 | 20.131 | 1558.6 | 268.1 | c |
| 5141 | 0.30468 | 71 | 19.86 | 20.44 | 0.58 | 20.247 | 1828.8 | 410.5 | c |
| 5155 | 0.56751 | 71 | 19.49 | 20.51 | 1.02 | 20.164 | 1645.6 | 476.0 | ab |
| 5330 | 0.39235 | 84 | 19.72 | 20.49 | 0.77 | 20.194 | 1701.6 | 586.8 | ab |
| 5343 | 0.54694 | 84 | 19.67 | 20.66 | 0.99 | 20.190 | 1964.0 | 621.7 | ab |
| 5344 | 0.64249 | 84 | 19.59 | 20.28 | 0.69 | 20.014 | 1988.6 | 627.0 | ab |
| 5354 | 0.67570 | 84 | 19.91 | 20.31 | 0.40 | 20.097 | 1530.3 | 712.4 | abB |
| 5359 | 0.67099 | 84 | 19.52 | 20.32 | 0.80 | 19.982 | 1528.6 | 732.6 | ab |
| 5364 | 0.28272 | 83 | 19.77 | 20.14 | 0.37 | 19.943 | 1898.5 | 744.6 | c |
| 5375 | 0.40155 | 84 | 19.77 | 20.28 | 0.51 | 19.990 | 1802.6 | 819.3 | c |
| 5376 | 0.38950 | 80 | 19.87 | 20.35 | 0.48 | 20.057 | 1623.3 | 820.7 | c |
| 5382 | 0.59593 | 84 | 19.76 | 20.58 | 0.81 | 20.164 | 1974.9 | 860.8 | ab |
| 5384 | 0.33483 | 84 | 19.96 | 20.43 | 0.47 | 20.091 | 1661.2 | 879.8 | c |
| 5390 | 0.65970 | 84 | 19.44 | 20.41 | 0.97 | 20.055 | 1575.5 | 910.7 | ab |

Table 1: - concluded

| ID | P(days) | Nv | Vmax | Vmin | Av | < V > | X | Y | type |
|------|---------|----|-------|-------|------|--------|--------|--------|------|
| 5393 | 0.32275 | 84 | 19.82 | 20.34 | 0.52 | 20.039 | 1677.9 | 916.6 | c |
| 5397 | 0.61731 | 84 | 19.61 | 20.47 | 0.85 | 20.137 | 1964.9 | 918.2 | ab |
| 5400 | 0.32892 | 84 | 19.99 | 20.53 | 0.55 | 20.217 | 1574.0 | 943.0 | c |
| 5401 | 0.50380 | 84 | 19.72 | 20.45 | 0.74 | 20.182 | 1558.4 | 947.8 | ab |
| 5492 | 0.52879 | 84 | 19.59 | 20.53 | 0.94 | 20.247 | 1875.2 | 949.9 | ab |
| 5496 | 0.52504 | 84 | 19.86 | 20.56 | 0.70 | 20.314 | 1565.6 | 961.6 | ab |
| 5710 | 0.35581 | 80 | 19.91 | 20.40 | 0.49 | 20.139 | 1879.8 | 1035.8 | c |
| 5714 | 0.29291 | 81 | 19.95 | 20.51 | 0.56 | 20.207 | 1633.4 | 1054.8 | c |
| 5721 | 0.35831 | 81 | 19.93 | 20.39 | 0.47 | 20.220 | 1722.5 | 1103.4 | c |
| 5723 | 0.56602 | 81 | 19.88 | 20.52 | 0.64 | 20.246 | 1523.9 | 1107.0 | ab |
| 5724 | 0.49852 | 80 | 19.71 | 20.56 | 0.85 | 20.344 | 1767.7 | 1107.4 | ab |
| 5728 | 0.25120 | 81 | 19.91 | 20.43 | 0.51 | 20.154 | 1637.3 | 1139.9 | c |
| 5730 | 0.35188 | 81 | 20.05 | 20.41 | 0.36 | 20.178 | 1785.8 | 1148.5 | c |
| 5747 | 0.55986 | 81 | 19.55 | 20.42 | 0.87 | 20.131 | 1548.1 | 1323.3 | ab |
| 5751 | 0.39732 | 81 | 19.84 | 20.36 | 0.51 | 20.070 | 1664.8 | 1361.6 | c |
| 5773 | 0.50878 | 81 | 19.56 | 20.63 | 1.07 | 20.207 | 1652.6 | 1011.9 | ab |
| 5778 | 0.61096 | 81 | 19.74 | 20.45 | 0.71 | 20.197 | 1889.1 | 1022.3 | ab |
| 5802 | 0.51458 | 81 | 19.47 | 20.67 | 1.20 | 20.378 | 1592.4 | 1167.9 | ab |
| 5828 | 0.53714 | 81 | 19.54 | 20.58 | 1.04 | 20.277 | 1816.4 | 1377.0 | ab |
| 5845 | 0.36134 | 81 | 19.85 | 20.38 | 0.52 | 20.138 | 1853.1 | 1473.3 | cB |
| 6031 | 0.24634 | 82 | 19.90 | 20.39 | 0.49 | 20.093 | 1685.4 | 1634.8 | c |
| 6032 | 0.50902 | 83 | 19.64 | 20.46 | 0.82 | 20.169 | 1753.3 | 1635.5 | ab |
| 6034 | 0.60893 | 83 | 19.69 | 20.29 | 0.60 | 20.034 | 1817.9 | 1660.0 | ab |
| 6048 | 0.62679 | 83 | 19.89 | 20.43 | 0.54 | 20.157 | 1939.8 | 1847.3 | ab |
| 6050 | 0.30517 | 83 | 19.89 | 20.38 | 0.50 | 20.131 | 1546.0 | 1920.6 | c |
| 6085 | 0.36153 | 83 | 19.93 | 20.40 | 0.47 | 20.147 | 1597.4 | 1807.9 | c |

Table 2: Equatorial coordinates of variables from the central part of the Sculptor galaxy. Coordinates of 3 anomalous cepheids and 2 variable red giants are given at the end of table.

| ID | RA(2000) deg | Dec(2000) deg | ID | RA(2000) deg | Dec(2000) deg | ID | RA(2000) deg | Dec(2000) deg |
|----|----------|-----------|----|----------|-----------|----|----------|-----------|
| 33 | 14.92432 | -33.83240 | 1491 | 14.98180 | -33.82260 | 2562 | 15.02418 | -33.70730 |
| 37 | 14.93299 | -33.82500 | 1519 | 15.00576 | -33.80140 | 2566 | 14.97853 | -33.70970 |
| 38 | 14.96458 | -33.82000 | 1546 | 14.99830 | -33.78680 | 2575 | 14.97613 | -33.70690 |
| 54 | 14.93999 | -33.79570 | 1553 | 15.04240 | -33.77860 | 2606 | 14.99543 | -33.69390 |
| 59 | 14.96322 | -33.78990 | 1555 | 14.99009 | -33.78160 | 2627 | 14.96712 | -33.68650 |
| 71 | 14.91777 | -33.78690 | 1558 | 14.97375 | -33.78110 | 2639 | 14.99691 | -33.67910 |
| 116 | 14.97255 | -33.79850 | 1566 | 14.99146 | -33.77700 | 2689 | 14.99448 | -33.65190 |
| 321 | 14.91151 | -33.76250 | 1823 | 14.97383 | -33.73860 | 2699 | 14.98388 | -33.65060 |
| 356 | 14.90564 | -33.73870 | 1824 | 14.98155 | -33.73710 | 2991 | 14.99496 | -33.64630 |
| 357 | 14.92735 | -33.73690 | 1830 | 15.01858 | -33.73350 | 3004 | 14.98054 | -33.63840 |
| 361 | 14.91075 | -33.73360 | 1838 | 15.00214 | -33.72870 | 3009 | 15.01450 | -33.63310 |
| 366 | 14.93720 | -33.72890 | 1873 | 14.98328 | -33.71600 | 3016 | 15.00302 | -33.63220 |
| 368 | 14.92808 | -33.72940 | 1874 | 14.99275 | -33.71540 | 3019 | 14.98840 | -33.63180 |
| 377 | 14.96484 | -33.72080 | 1875 | 15.02698 | -33.71310 | 3024 | 14.96215 | -33.62770 |
| 385 | 14.90586 | -33.71970 | 1877 | 14.97852 | -33.71540 | 3026 | 15.01235 | -33.62330 |
| 403 | 14.90195 | -33.76820 | 1890 | 15.00592 | -33.76470 | 3039 | 15.02401 | -33.60620 |
| 406 | 14.96991 | -33.76240 | 1899 | 14.98055 | -33.76910 | 3043 | 15.02532 | -33.60430 |
| 411 | 14.93304 | -33.76310 | 1910 | 15.02182 | -33.76390 | 3044 | 14.97222 | -33.60580 |
| 416 | 14.92731 | -33.76240 | 1914 | 15.01456 | -33.76340 | 3104 | 15.01257 | -33.61650 |
| 439 | 14.93767 | -33.74220 | 1926 | 14.96991 | -33.76240 | 3113 | 14.98833 | -33.61170 |
| 462 | 14.90621 | -33.73410 | 1930 | 15.02574 | -33.75580 | 3125 | 15.02585 | -33.59860 |
| 463 | 14.92079 | -33.73320 | 1932 | 14.97745 | -33.75870 | 3126 | 15.01938 | -33.59850 |
| 493 | 14.90354 | -33.71900 | 1940 | 14.98223 | -33.75510 | 3143 | 15.01764 | -33.58590 |
| 737 | 14.90769 | -33.71500 | 1941 | 15.00118 | -33.75400 | 3318 | 15.09179 | -33.82100 |
| 753 | 14.93654 | -33.70360 | 1943 | 15.03828 | -33.75030 | 3319 | 15.11376 | -33.81950 |
| 763 | 14.92973 | -33.69650 | 1997 | 15.02578 | -33.72920 | 3320 | 15.09673 | -33.81640 |
| 771 | 14.90183 | -33.69460 | 2004 | 15.02232 | -33.72730 | 3345 | 15.04600 | -33.79500 |
| 782 | 14.90648 | -33.68890 | 2012 | 15.02552 | -33.72540 | 3346 | 15.08145 | -33.79150 |
| 786 | 14.93024 | -33.68370 | 2021 | 15.00775 | -33.72410 | 3365 | 15.05007 | -33.78540 |
| 793 | 14.92805 | -33.67760 | 2048 | 15.00851 | -33.71150 | 3397 | 15.09272 | -33.79580 |
| 803 | 14.94209 | -33.66870 | 2058 | 15.03372 | -33.70740 | 3410 | 15.08215 | -33.81740 |
| 811 | 14.95219 | -33.66280 | 2059 | 14.99844 | -33.70960 | 3413 | 15.08767 | -33.81460 |
| 853 | 14.90020 | -33.70210 | 2410 | 15.00678 | -33.70740 | 3468 | 15.06128 | -33.77480 |
| 860 | 14.90696 | -33.69640 | 2421 | 14.99032 | -33.70290 | 3710 | 15.07710 | -33.75970 |
| 1132 | 14.93776 | -33.64910 | 2422 | 15.03073 | -33.70020 | 3760 | 15.05825 | -33.73000 |
| 1139 | 14.95097 | -33.63840 | 2423 | 15.03667 | -33.69950 | 3761 | 15.08222 | -33.72720 |
| 1142 | 14.93833 | -33.63660 | 2424 | 14.97123 | -33.70340 | 3763 | 15.10665 | -33.72490 |
| 1149 | 14.92897 | -33.63090 | 2425 | 14.98572 | -33.70180 | 3777 | 15.07337 | -33.71920 |
| 1164 | 14.90931 | -33.61670 | 2450 | 14.98841 | -33.69240 | 3801 | 15.07928 | -33.70810 |
| 1168 | 14.92255 | -33.60710 | 2455 | 14.96500 | -33.69000 | 3810 | 15.05889 | -33.70640 |
| 1178 | 14.89258 | -33.59340 | 2458 | 15.01054 | -33.68580 | 3827 | 15.05972 | -33.75960 |
| 1204 | 14.90172 | -33.64160 | 2467 | 15.03063 | -33.68070 | 3832 | 15.08860 | -33.75540 |
| 1256 | 14.89872 | -33.59970 | 2470 | 14.99779 | -33.68190 | 3834 | 15.10492 | -33.75360 |
| 1261 | 14.94661 | -33.59230 | 2471 | 15.00894 | -33.68080 | 3862 | 15.09263 | -33.74140 |
| 1411 | 14.99472 | -33.82210 | 2482 | 14.96870 | -33.68130 | 3888 | 15.09075 | -33.72890 |
| 1424 | 15.04201 | -33.80870 | 2502 | 14.96484 | -33.67140 | 3907 | 15.05722 | -33.72330 |
| 1439 | 15.04600 | -33.79500 | 2528 | 15.01420 | -33.66130 | 3916 | 15.04278 | -33.72040 |
| 1446 | 15.03224 | -33.79090 | 2545 | 15.00620 | -33.65140 | 3931 | 15.03899 | -33.71260 |
| 1457 | 15.00387 | -33.78500 | 2552 | 14.98672 | -33.69880 | 3934 | 15.05840 | -33.70880 |
| 1462 | 15.02212 | -33.77960 | 2555 | 14.98551 | -33.67890 | 3938 | 15.08497 | -33.70530 |
| 1470 | 15.02172 | -33.77460 | 2558 | 15.03372 | -33.70740 | 3941 | 15.10270 | -33.70320 |
| 1482 | 15.03707 | -33.77180 | 2559 | 14.99844 | -33.70960 | 4233 | 15.03667 | -33.69950 |

Table 2: - concluded

| ID | RA(2000) deg | Dec(2000) deg | ID | RA(2000) deg | Dec(2000) deg | ID | RA(2000) deg | Dec(2000) deg |
|---|---|---|---|---|---|---|---|---|
| 4235 | 15.10594 | -33.69290 | 5068 | 15.16510 | -33.76340 | 5721 | 15.14097 | -33.68780 |
| 4263 | 15.07828 | -33.67790 | 5081 | 15.12769 | -33.77430 | 5723 | 15.11217 | -33.68930 |
| 4272 | 15.05915 | -33.67680 | 5085 | 15.19328 | -33.81490 | 5724 | 15.14747 | -33.68680 |
| 4277 | 15.08035 | -33.67180 | 5105 | 15.16110 | -33.80110 | 5728 | 15.12821 | -33.68420 |
| 4291 | 15.06657 | -33.66590 | 5123 | 15.12700 | -33.79000 | 5730 | 15.14962 | -33.68170 |
| 4308 | 15.03822 | -33.65980 | 5141 | 15.16452 | -33.77020 | 5747 | 15.11315 | -33.66300 |
| 4309 | 15.08283 | -33.65640 | 5155 | 15.13718 | -33.76410 | 5751 | 15.12959 | -33.65730 |
| 4313 | 15.09362 | -33.65130 | 5330 | 15.14401 | -33.75020 | 5773 | 15.13192 | -33.69940 |
| 4353 | 15.03838 | -33.69450 | 5343 | 15.18163 | -33.74350 | 5778 | 15.16605 | -33.69590 |
| 4385 | 15.04945 | -33.67690 | 5344 | 15.18513 | -33.74260 | 5802 | 15.12139 | -33.68130 |
| 4686 | 15.09342 | -33.64120 | 5354 | 15.11771 | -33.73670 | 5828 | 15.15135 | -33.65390 |
| 4689 | 15.06677 | -33.64140 | 5359 | 15.11723 | -33.73430 | 5845 | 15.15554 | -33.64200 |
| 4747 | 15.09084 | -33.64160 | 5364 | 15.17069 | -33.72930 | 6031 | 15.12937 | -33.62420 |
| 4780 | 15.05741 | -33.62590 | 5375 | 15.15591 | -33.72120 | 6032 | 15.13918 | -33.62350 |
| 4785 | 15.07185 | -33.62000 | 5376 | 15.12991 | -33.72280 | 6034 | 15.14825 | -33.61990 |
| 4786 | 15.08642 | -33.61900 | 5382 | 15.18039 | -33.71450 | 6048 | 15.16367 | -33.59620 |
| 4793 | 15.08267 | -33.61270 | 5384 | 15.13471 | -33.71530 | 6050 | 15.10588 | -33.59130 |
| 4812 | 15.07712 | -33.59840 | 5390 | 15.12194 | -33.71240 | 6085 | 15.11462 | -33.60430 |
| 4824 | 15.05913 | -33.58620 | 5393 | 15.13670 | -33.71070 | 3302 | 15.06876 | -33.79940 |
| 5000 | 15.17154 | -33.81570 | 5397 | 15.17826 | -33.70770 | 734 | 14.89081 | -33.65870 |
| 5011 | 15.12622 | -33.80470 | 5400 | 15.12135 | -33.70850 | 5689 | 15.12568 | -33.69510 |
| 5015 | 15.14345 | -33.79860 | 5401 | 15.11903 | -33.70810 | 274 | 14.89677 | -33.73590 |
| 5030 | 15.12805 | -33.78770 | 5492 | 15.16490 | -33.70480 | 687 | 14.94274 | -33.68980 |
| 5032 | 15.14285 | -33.78460 | 5496 | 15.11992 | -33.70630 | | | |
| 5049 | 15.16763 | -33.77230 | 5710 | 15.16454 | -33.69440 | | | |
| 5065 | 15.14519 | -33.76550 | 5714 | 15.12864 | -33.69450 | | | |

Table 3: Light curve parameters and rectangular coordinates of Sculptor anomalous cepheids and variable red giants. Coordinates X and Y are expressed in pixels (1 pixel equals 0.435").

| ID | P(days) | Nv | Vmax | Vmin | Av | $<V>$ | X | Y |
|---|---|---|---|---|---|---|---|---|
| 3302 | 1.34607 | 75 | 18.10 | 18.91 | 0.80 | 18.545 | 1153.2 | 222.4 |
| 734 | 1.15776 | 71 | 18.53 | 19.08 | 0.55 | 18.864 | 22.7 | 1480.7 |
| 5689 | 0.85541 | 81 | 18.65 | 19.35 | 0.70 | 19.135 | 1612.7 | 1051.3 |
| 274 | | 69 | 17.23 | 17.12 | 0.11 | 17.17 | 13.8 | 838.5 |
| 687 | | 83 | 17.09 | 17.20 | 0.11 | 17.15 | 359.7 | 1194.8 |

Table 4: Properties of RR Lyr variables in four dwarf spheroidal galaxies. $N_{RR}$ is the total number of RR Lyr variables with known periods. $n_{ab}$, $n_c$ and $n_d$ are relative frequencies of occurrence for RRab, RRc and RRd subtypes. Average periods for RRab and RRc variables are given in columns 6 and 7, respectively.

| Object | $N_{RR}$ | $n_{ab}$ | $n_c$ | $n_d$ | $<P_{ab}>$ | $<P_c>$ | [Fe/H] |
|---|---|---|---|---|---|---|---|
| Carina | 58 | 0.845 | 0.155 | 0.000 | 0.620 | 0.366 | -2.2 |
| Draco | 131 | 0.878 | 0.046 | 0.076 | 0.614 | 0.351 | -2.4 |
| Sculptor | 221 | 0.602 | 0.368 | 0.013 | 0.586 | 0.340 | -1.9 |
| Ursa Minor | 82 | 0.573 | 0.427 | 0.000 | 0.638 | 0.375 | -2.0 |